# INTERACTIVE AUDIO-TACTILE MAPS FOR VISUALLY IMPAIRED PEOPLE

Anke Brock
Inria Bordeaux, France
anke.brock@inria.fr

Christophe Jouffrais
CNRS & Université de Toulouse; IRIT; France
christophe.jouffrais@irit.fr

## Introduction

Visually impaired people face important challenges related to orientation and mobility. Indeed, 56% of visually impaired people in France declared having problems concerning autonomous mobility [10]. These problems often mean that visually impaired people travel less, which influences their personal and professional life and can lead to exclusion from society [28]. Therefore this issue presents a social challenge as well as an important research area. Accessible geographic maps are helpful for acquiring knowledge about a city's or neighborhood's configuration, as well as selecting a route to reach a destination. Traditionally, raised-line paper maps with braille text have been used. These maps have proved to be efficient for the acquisition of spatial knowledge by visually impaired people. Yet, these maps possess significant limitations [37]. For instance, due to the specificities of the tactile sense only a limited amount of information can be displayed on a single map, which dramatically increases the number of maps that are needed. For the same reason, it is difficult to represent specific information such as distances. Finally, braille labels are used for textual descriptions but only a small percentage of the visually impaired population reads braille. In France 15% of blind people are braille readers and only 10% can read and write [10]. In the United States, fewer than 10% of the legally blind people are braille readers and only 10% of blind children actually learn braille [24].

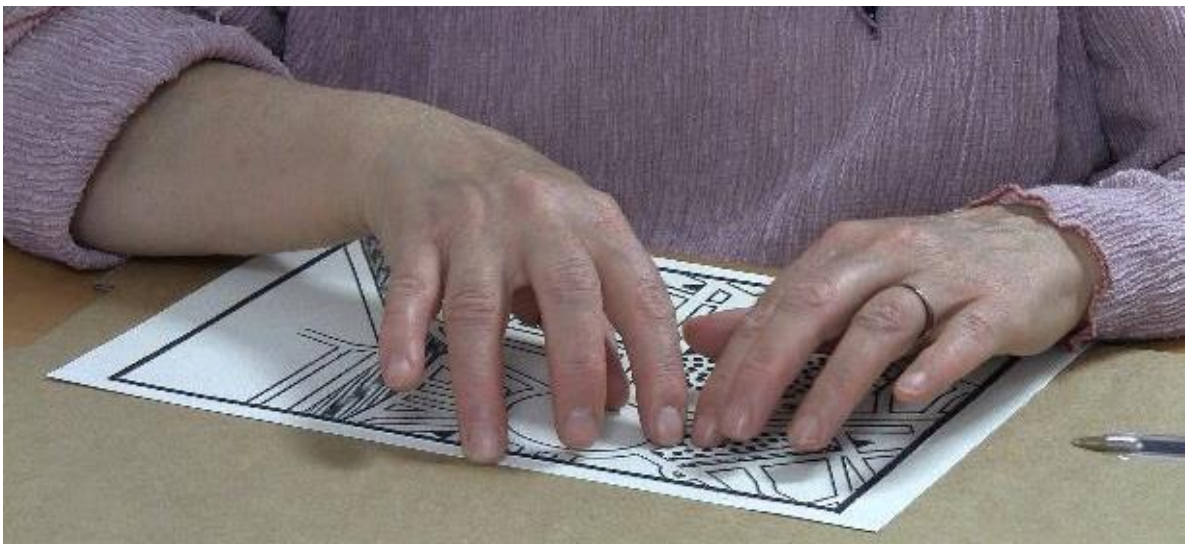

Figure 1: A visually impaired person reading a tactile map

Recent technological advances have enabled the design of interactive maps with the aim to overcome these limitations. Indeed, interactive maps have the potential to provide a broad spectrum of the population with spatial knowledge, irrespective of age, impairment, skill level, or other factors [25]. To this regard, they might be an efficient means for providing visually impaired people with access to geospatial information. In this paper we give an overview of our research on making geographic maps accessible to visually impaired people.

## Related Work

In order to classify existing accessible interactive maps, we performed an exhaustive search through scientific databases (ACM Digital Library, SpringerLink, IEEE Explorer, and Google Scholar). We found 43 articles that were published between 1988 and 2013 that matched our inclusion criteria. First, we only considered interactive maps (and not navigation systems) and only those maps that were specifically designed for visually impaired people. Second, we only included publications in journals or peer-reviewed conferences. Third, for identic prototypes we only considered one publication. We presented the classification of this map corpus in [4,5] in detail.

To sum up, the design of accessible interactive maps varied in different aspects, such as content, devices, or input and output interaction techniques. We observed that most accessible interactive map prototypes rely on touch input, and some systems use both touch and audio (speech recognition) input [1,12,15,16,36]. All systems rely on audio output, except [20] which was entirely based on the tactile modality. In Figure 2 we propose a classification according to the devices used in the prototype. We classified the devices in four categories: haptic devices, tactile actuator devices, touch-sensitive devices and other.

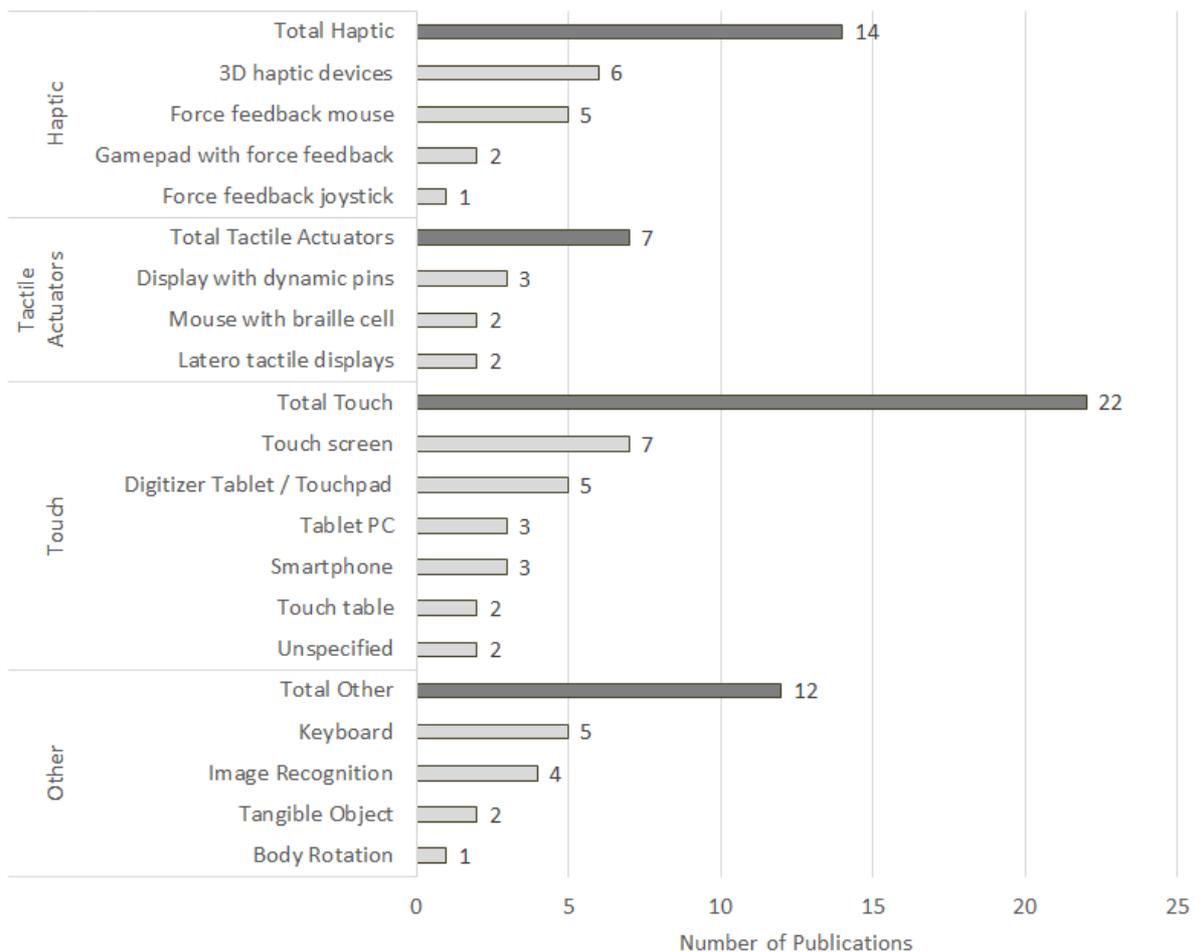

Figure 2: Classification of accessible interactive map prototypes by type of device.

Within the "haptic devices" category, we included devices that provide force feedback. This means that they mechanically produce a force that is perceived as a kinesthetic sensation by the user. Examples for prototypes using these type of technology are presented in [13,26,34,36].

"Tactile actuators" dynamically provide a cutaneous stimulation on the user's skin. Many of those systems use needles or pins that are mechanically raised. Many visually impaired people are familiar with this kind of stimulation, as they often use dynamic braille displays that are based on a similar principle [2]. A few interactive map prototypes are based on large displays with actuated pins (see for instance [41]). However, as the production of large displays is expensive, smaller displays have been used, e.g. in the size of braille cells (see [38] as an example). Another solution, relies on tactile feedback which is produced by laterally stretching the skin of the finger [29].

Within the "touch-sensitive" devices category we gathered different technologies that locate touch inputs (mono and multi-touch, using bare fingers or pens). Touch-based surfaces per se do not provide tactile feedback and thus are usually combined with audio feedback [1,16]. In some accessible map prototypes, the touch-based device was combined with a raised-line map overlay [19,22,27,39] or with vibro-tactile stimulation [33,40] in order to ease the exploration with a supplementary tactile information.

The last category includes all the prototypes with a different technology. For instance, some prototypes integrate keyboards as supplementary input device for entering textual information [1,19,26]. Other prototypes are based on image recognition (e.g. [15]) allowing to localize the hand relative to the map and provide audio feedback consequently. The user thus interacts with the map as if it was based on a touch-sensitive surface. Few projects investigated tangible interaction, i.e. interaction through physical objects [23,32]. Finally, one project [23] used the rotation of the user's own body for controlling the map orientation.

## Design and evaluation of an audio-tactile map prototype

### Design of an audio-tactile map prototype

Based on the survey of the related work and our own observations of visually impaired users during mobility and orientation lessons [see 14], we developed an accessible interactive map prototype. The interactive map prototype was composed of a raised-line map overlay placed over a multi-touch screen (see Figure 3), a computer connected to the screen and loudspeakers for speech output. Users could explore the raised-line map on top of the screen with both hands, i.e. ten fingers, exactly the same way that they would explore a tactile paper map. Instead of reading a braille legend, they could obtain the names of streets and buildings by double-tapping on the map elements. The development of this map prototype consisted in three processes [see 6]: 1) drawing and printing the raised-line paper maps, 2) choosing the multi-touch technology, and 3) designing and implementing non-visual interaction methods.

We chose to use microcapsule paper for printing the raised-line maps because it is the easiest production method and has been successfully used in previous audio-tactile map prototypes [22,39]. Another important argument was that this paper is slim, which is advantageous to detect touch input through the paper map. Maps were designed with the Open Source Inkscape software using SVG (Scalable Vector Graphics) format. There are no strict rules on how to design tactile maps and those maps use different symbols and textures. However different guidelines exist [3,31,37] which we respected when designing our prototype (Figure 3, [8]).

Concerning the choice of a suitable multi-touch technology we identified several criteria. The most important one was reliable touch detection through a paper overlay. After diverse tests, we chose the 3M Inc. multi-touch screen model M2256PW. The capacitive projected technology ensured pre-served responsiveness through the paper overlay. Furthermore, the dimensions of the screen (slightly larger than A3 format) were well adapted for representing a city neighborhood.

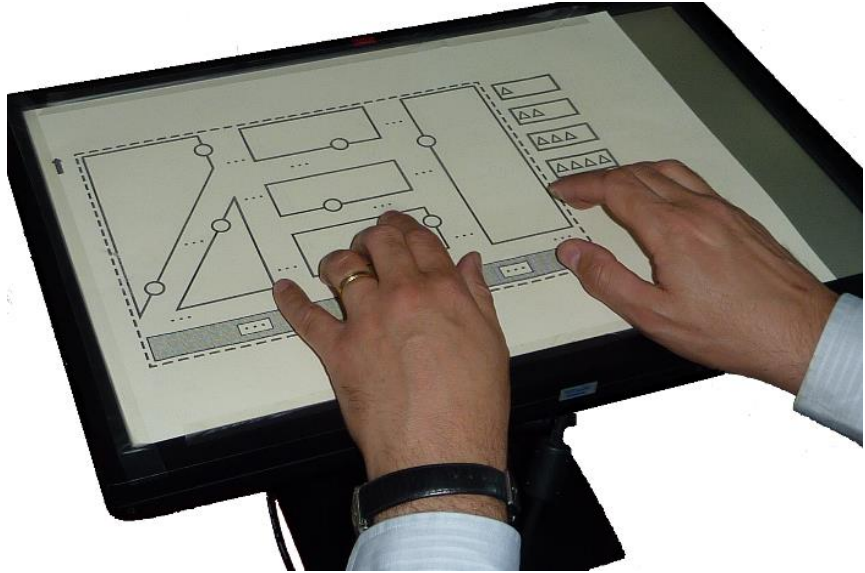

Figure 3: Interactive audio-tactile map prototype composed of a multi-touch screen with raised-line overlay

With regard to interaction techniques, the prototype provided audio-tactile output. The raised lines of the tactile map were the first available sensory cues. For audio output, we used the Realspeak SAPI 4.0 French text-to-speech synthesis (voice "Sophie") which possesses a good intelligibility and user appreciation [11]. It has been observed in several studies that simple taps are frequently generated during exploratory hand movements and thus lead to unintended feedbacks [21]. Consequently, we implemented a double-tap input.

## Evaluating the usability of the audio-tactile map

Prior to our project, the usability of accessible interactive maps had never been compared to the usability of raised-line maps with braille text. Therefore, it was unknown whether interactive maps were worse or better solutions than traditional raised-line maps. To overcome this lack of knowledge, we conducted a systematic user study, comparing these two different map types for visually impaired people [8]. Our general hypothesis was that an interactive map (IM) was more usable than a tactile paper map (PM) for providing blind people with spatial knowledge about a novel environment.

We designed a map representing a fictional city center with six streets, six buildings, six points of interest (e.g., museum, restaurant, and metro station), as well as a river (see Figure 3). A second map was created with the same map elements that were rotated and translated, so that both map contents were equivalent. The hotel as central point of interest was common for both maps. One of the maps contained braille labels and was accompanied by a legend explaining those labels (PM) whereas the other map did not have any braille label but was interactive and provided audio feedback (IM). We ensured lexical equivalence of the names of streets and POIs, and we made pretests with a visually impaired user to ensure that the maps were readable, and that they were not too easy or too difficult to memorize.

Both maps were tested by twenty-four blind participants recruited from different associations, through a local radio broadcast for visually impaired people as well as by word-of-mouth. The experimental protocol included a short- and a long-term study that were each composed by two sessions. In this paper we only report about the short-term study (the long-term study is described in [8]). The two sessions took place in a dedicated experimental environment in the IRIT research laboratory in Toulouse, France. During the first session, the subjects first explored a simple map during a familiarization phase, and then answered a questionnaire about personal characteristics. Then, they were asked to explore and learn the first map (either IM or PM depending on the group) with both accuracy and time constraints ("as quickly and as accurate as possible"). Participants were informed that they would have to answer questions afterwards without having access to the map. In order to motivate them to memorize the map, users were asked to prepare holidays in an unknown city and we invited them to memorize the map in order to fully enjoy the trip. Users were free to explore until they considered that they had completely memorized the map. When they stopped, we measured the learning time and removed the map. Subjects then answered questions evaluating the three types of spatial knowledge (landmark, route, survey). The second session took place one week later and started with a familiarization phase followed by an interview on spatial abilities. Participants then explored the second map type (either PM or IM) and responded to the questions on spatial knowledge. We finally assessed their satisfaction regarding the two maps.

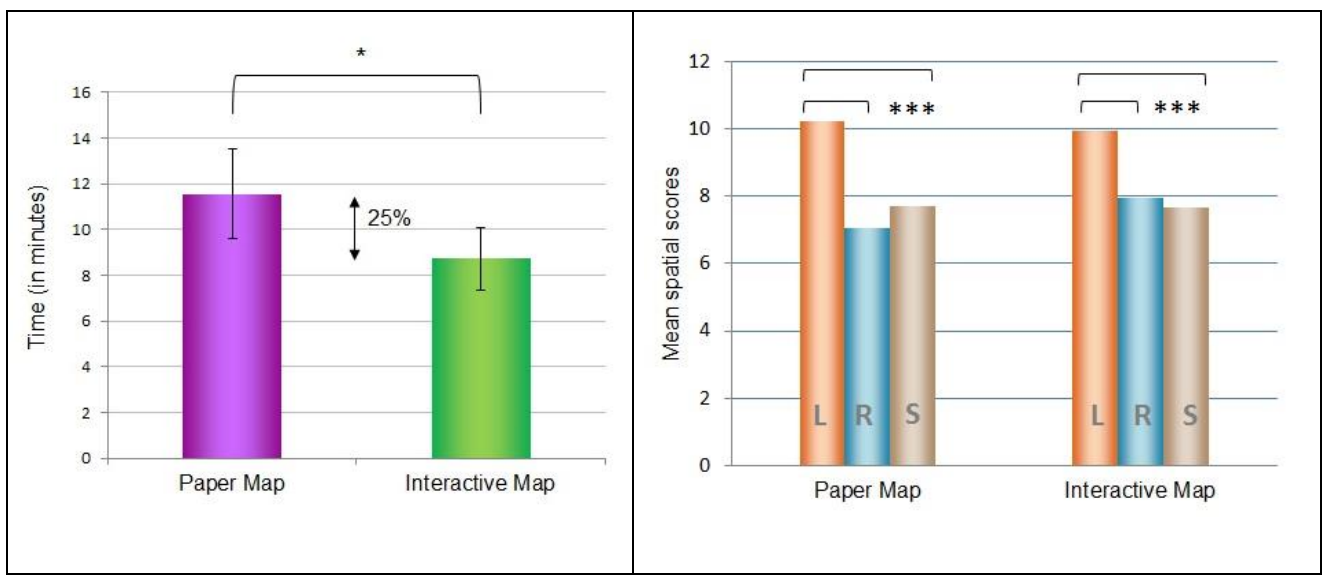

Figure 4: Experimental Results
(a) Learning Time (mean values measured in minutes) for the paper map (left) as compared to the interactive map (right). The Learning Time for the interactive map was significantly lower than for the paper map (lower is better).
(b) Mean spatial scores for responses to landmark (orange), route (blue) and survey (brown) questions (paper map left, interactive map right). Mean scores for the landmark tasks were significantly higher than those for the route and survey tasks (higher is better). There was no significant difference between R and S questions. * p < .05, *** p< .001

We made the assumptions that: 1/ exploration duration (corresponding to the learning time) reflects efficiency; 2/ the quality of spatial learning (measured as spatial scores) reflects effectiveness; 3/ map preference reflects user satisfaction. Learning time was significantly shorter for the interactive map than for the paper map. As shown in Figure 4(a), using a 2*2 repeated measures ANOVA (F(1,22) = 4.59, p = .04), Learning Time appeared to be significantly shorter for the interactive map (*M* = 8.71, *SD* = 3.36) than for the paper map (*M* = 11.54, *SD* = 4.88). Regarding spatial learning, there was no significant difference between both map types. However, significant differences were observed when comparing scores for L, R, and S questions (Figure 4 b). Pairwise Wilcoxon rank sum tests with Bonferroni correction (alpha level = .017) revealed that the difference between L (*M* = 10.1, *SD* = 2.0) and R (*M* = 7.5, *SD* = 2.9) was significant (N = 45, Z = 5.20, p < .001) as well as the difference between L and S (*M* = 7.7, *SD* = 2.7) questions (N = 43, Z = 5.06, p < .001). There was no significant difference between R and S questions (N = 41, Z = 0.41, p = .68). Finally, when asked which map they preferred, more users answered in favor of the interactive map: of a total of 24 users, 17 preferred the interactive map, six users preferred the paper map, and one had no preference. Furthermore, as reported in [8], we also observed correlations between the observed variables and personal characteristics. As two examples, people who are frequently using new technologies (smartphones, computers) needed more time for reading the paper map with braille text, and early blind people and those who were better braille readers experienced a higher satisfaction towards reading the paper map. To roughly sum up, this study confirmed that visually impaired people are able to memorize and mentally manipulate both route and survey spatial knowledge. In addition, it demonstrated that interactive audio-tactile maps are more usable than regular tactile maps with braille text for blind users.

### Non-visual gestural interaction for geographic maps

The non-visual interaction techniques implemented in our experimental prototype as described above were quite simple. The main interaction was a double tap that allowed receiving vocal feedback about names of streets, buildings, etc. Following this study we aimed at including supplementary information—such as opening hours, entry prices, distances, etc.—in the map prototype. One possibility to make touch screens accessible without sight is using gestural interaction [17,21]. However, gestural interaction had never been combined with a raised-line map overlay. We implemented basic gestural interaction techniques provided by the MT4J API [18], an open-source and cross-platform framework for developing multi-touch applications [6]. Among these we selected a lasso gesture (i.e., circling around a map element without lifting the finger) to retrieve information associated with points of interest. Additionally, we implemented a tap and hold gesture. In this case, the user had to tap on a map element and maintain the finger pressed until a beep sound confirmed the activation. When tapping on a second map element, another beep confirmed the activation, and the distance between these elements was announced.

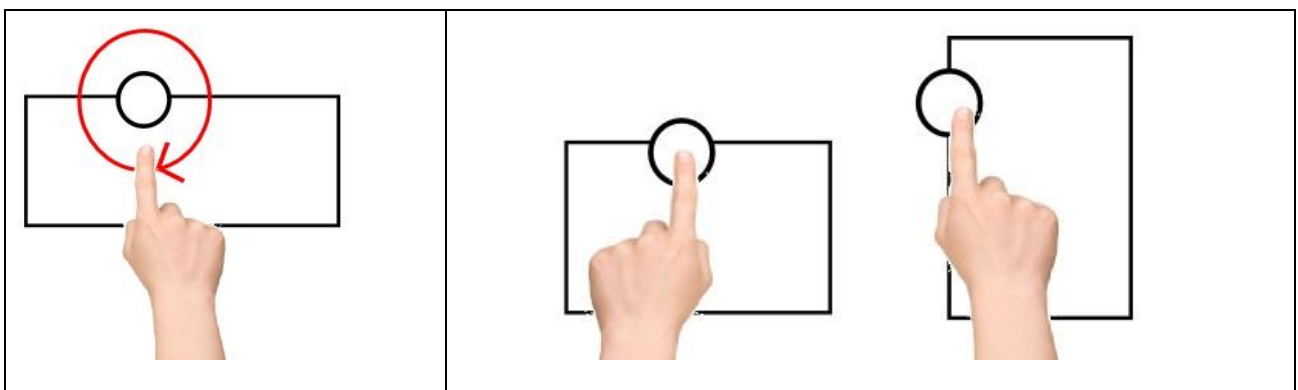

Figure 5: Gestural Interaction (a) Lasso, (b) Tap and Hold

### Perspectives

In the continuity of the research presented in this paper, the AccessiMap project (http://www.irit.fr/accessimap) has the objective to exploit the availability of Open Data (e.g.

OpenStreetMap) and make them accessible using tangible interaction. The consortium consists of two research centers, a company, and a specialized center for visually impaired people. It pursues several technical and scientific objectives including: designing a web-based editor allowing graphics transcribers to quickly elaborate tactile graphics relying on Open Data; designing a prototype of a collaborative tangible tabletop allowing the evaluation of non-visual interactions for graphical data exploration; assessing mental spatial representations of visually impaired users after exploration; improving the overall accessibility of Open Data; and launching cheap or free accessible apps for tablets and smartphones.

## Designing with and for visually impaired users

An important principle in HCI research is to include users throughout the whole design cycle in order to ensure that the developed technologies meet the users' needs. This can be done through the use of participatory design or codesign methods [35]. This principle is also very important—if not even more important—when designing for people with special needs. Indeed, designers or researchers without impairments cannot easily design adapted assistive technologies. Basing the development of assistive technology on the emergence of new technologies and not on users' needs, leads to a high abandoning rate [30].

Our research with visually impaired people relied on participatory design methods for all design phases: analysis, ideation, prototyping and evaluation. Through a close collaboration with the Institute of Young Blinds in Toulouse (CESDV-IJA), we have been able to meet a large number of visually impaired people as well as locomotion trainers and specialized teachers. As most design methods make use of the visual sense (for instance, sharing of ideas during a brainstorming session by writing them on a whiteboard), we have experienced the need to adapt existing methods accordingly when working with blind people [9].

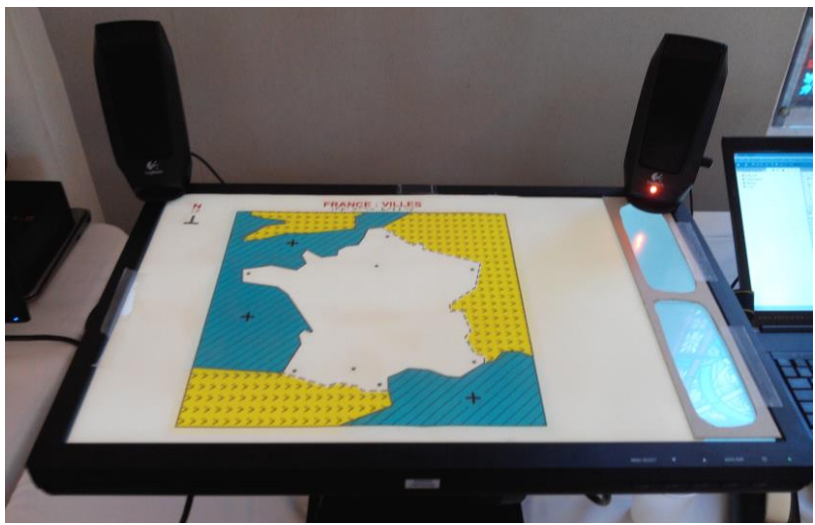

Figure 6: Adapted Version of the interactive map prototypes as it is used in the Institute of Young Blinds in Toulouse (CESDV-IJA) for teaching geography to visually impaired children.

The close collaboration with the Institute of Young Blinds has enabled us to move our prototypes from the lab to the field. Thus, an extended version of the above mentioned prototype is currently used in classrooms for teaching geography to visually impaired children. Figure 6 illustrates a geography lesson where France is represented with surrounding countries and seas. In this map, many levels of information—such as local dialects and music—are associated to each point of interest. The user can navigate between the different levels using the menus on the right.

## Acknowledgments

Most importantly, we want to thank all visually impaired users who, over the duration of several years, have participated in design sessions and user studies. We also thank Claude Griet and the other teachers and trainers from CESDV-IJA and LACII, Toulouse, France, for their help and availability. Furthermore, we thank Delphine Picard, Bernard Oriola, Philippe Truillet, as well as several master students who have participated in our research. Gregoire Denis and Mustapha Ennadif have adapted the prototype for use in the classroom. This work was partially supported by the French National Research Agency (project NAVIG n°ANR-08-TECS-011).

**About the Authors:**

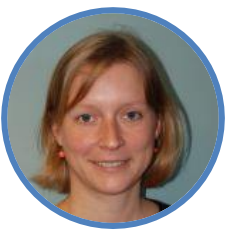

**Dr. Anke Brock** is a research scientist in HCI at Inria Bordeaux, France. She obtained a PhD (2013) and Master (2010) in Human-Computer Interaction from the University of Toulouse, France, and an engineering diploma in Information Technology from Baden-Wuerttemberg Cooperative State University (Stuttgart, Germany, 2004). Her research interests include assistive technology for visually impaired people and advanced interaction techniques for geographic maps.

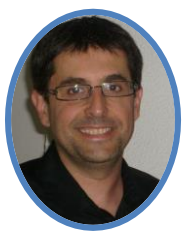

**Dr. Christophe Jouffrais** is with the IRIT Lab (UMR5505, CNRS & Univ of Toulouse) in Toulouse, FR. He is a CNRS researcher with a background in Cognitive Neuroscience, HCI and Assistive Technology. His current research focuses on perception, action and cognition without vision with an emphasis on non-visual Human-Computer Interaction, and Assistive Technologies for visually impaired people.